\documentclass[aps,preprint,nobibnotes,nofootinbib]{revtex4}

\usepackage{amsmath}

\begin{document}

\title{Derivation of Gravitational Self-Force}

\author{Samuel E. Gralla and Robert M. Wald \\ \it Enrico
Fermi Institute and Department of Physics \\ \it University of Chicago
\\ \it 5640 S.~Ellis Avenue, Chicago, IL~60637, USA}

\begin{abstract}
We analyze the issue of ``particle motion'' in general relativity in a systematic and rigorous way by considering a one-parameter family of metrics corresponding to having a body (or black hole) that is ``scaled down'' to zero size and mass in an appropriate manner. We prove that the limiting worldline of such a one-parameter family must be a geodesic of the background metric and obtain the leading order perturbative corrections, which include gravitational self-force, spin force, and geodesic deviation effects.  The status the MiSaTaQuWa equation is explained as a candidate ``self-consistent perturbative equation'' associated with our rigorous perturbative result.
\end{abstract}

\maketitle

\bigskip
\bigskip

It is of considerable interest to determine the motion of a body in
general relativity in the limit of small size, taking into account the
deviations from geodesic motion arising from gravitational self-force
effects. There is a general consensus that
the gravitational self-force is given by the ``MiSaTaQuWa equations'':
In the absence of incoming radiation, the motion is given by
\begin{equation}
u^\nu \nabla_\nu u^\mu = - \frac{1}{2} (g^{\mu \nu} + u^\mu u^\nu)(2\nabla_\sigma
h_{\nu \rho}^{\tiny \textrm{tail}}-\nabla_\nu h_{\rho \sigma}^{\tiny
\textrm{tail}})\big |_{z(\tau)}u^\rho u^\sigma \,\, ,
\end{equation}
\begin{equation}
h_{\mu \nu}^{\tiny \textrm{tail}}(x) = M
\int_{-\infty}^{\tau^-}\left(G^+_{\mu \nu \mu'
\nu'}-\frac{1}{2}g_{\mu \nu}G^{+ \ \rho}_{\ \rho \ \mu ' \nu
'}\right)\left(x,z(\tau ')\right)u^{\mu '}u^{\nu '} d\tau ' \,\, ,
\end{equation}
where $G_+$ is the retarded Green's function for the wave operator
$\nabla^\alpha \nabla_\alpha \tilde{h}_{\mu \nu} - 
2R^\alpha{}_{\mu \nu}{}^\beta \tilde{h}_{\alpha \beta}$. 
(Note that the $\tau^-$ limit of integration indicates that only the part
of $G_+$ interior to the light cone contributes to 
$h_{\mu \nu}^{\textrm{\tiny tail}}$.)
However, all derivations contain some unsatisfactory features.  This
is not surprising in view of the fact that, as noted in \cite{this-volume}, ``point particles'' do not make sense in nonlinear theories like general relativity!

\begin{itemize}

\item Derivations that treat the body as a point particle require
unjustified ``regularizations''.

\item Derivations using matched asymptotic expansions \cite{matched-expansions}
make a number of ad hoc and/or unjustified assumptions.

\item The axioms of the Quinn-Wald axiomatic approach \cite{quinn-wald} have not been
shown to follow from Einstein's equation.

\item All of the above derivations employ at some stage a ``phoney''
version of the linearized Einstein equation with a point particle
source, wherein the Lorenz gauge version of the linearized Einstein
equation is written down, but the Lorenz gauge condition is not imposed.

\end{itemize}

How should gravitational self-force be rigorously derived?
A precise formula for gravitational self-force can hold only in a
limit where the size, $R$, of the body goes to zero. Since
``point-particles'' do not make sense in general
relativity---collapse to a black hole would occur before a
point-particle limit could be taken---the mass, $M$, of the body must
also go to zero as $R \rightarrow 0$. In the limit as $R,M \rightarrow
0$, the worldtube of the body should approach a curve, $\gamma$, which
should be a geodesic of the ``background metric''. The self-force
should arise as the lowest order in $M$ correction to $\gamma$. In the following,
we shall describe an approach that we have recently taken to derive gravitational self-force
in this manner. Details can be found in \cite{gralla-wald}.

The discussion above suggests that we consider a one-parameter family of solutions to
Einstein's equation, $(g_{\mu\nu}(\lambda), T_{\mu\nu} (\lambda))$, with
$R(\lambda) \rightarrow 0$ and $M(\lambda) \rightarrow 0$ as $\lambda
\rightarrow 0$. But, what conditions should be imposed on
$(g_{\mu\nu}(\lambda), T_{\mu\nu} (\lambda))$ to ensure that it corresponds to
a body that is shrinking down to zero size, but is not undergoing wild
oscillations, drastically changing its shape, or doing other crazy
things as it does so?

As a very simple, explicit example of the kind of one-parameter family we seek, consider the Schwarzschild-deSitter metrics with $M = \lambda$,
\begin{equation}
ds^2 (\lambda) = -(1-\frac{2\lambda}{r} - Cr^2)dt^2 + (1-\frac{2\lambda}{r} -Cr^2)^{-1}dr^2 + 
r^2 d\Omega^2.
\end{equation}
If we take the limit as $\lambda \rightarrow 0$ at fixed coordinates
$(t,r,\theta,\phi)$ with $r>0$, it is easily seen that we obtain the deSitter
metric---with the deSitter spacetime worldline $\gamma$ defined by
$r=0$ corresponding to the location of the black hole ``before it
disappeared''.
However, there is also another limit that can be taken. At each time $t_0$, one 
can ``blow up''
the metric $g_{\mu\nu} (\lambda)$ by multiplying it by $\lambda^{-2}$,
i.e., define 
\begin{equation}
\bar{g}_{\mu\nu} (\lambda) \equiv \lambda^{-2}g_{\mu\nu}(\lambda).
\end{equation}
We correspondingly rescale the coordinates by defining
$\bar{r} = r/\lambda$, $\bar{t} = (t-t_0)/\lambda$. Then
\begin{equation}
d\bar{s}^2 (\lambda) = -(1-2/\bar{r} - \lambda^2C\bar{r}^2)d\bar{t}^2 + (1-2/\bar{r} - \lambda^2C\bar{r}^2)^{-1}d\bar{r}^2 + 
\bar{r}^2 d\Omega^2
\end{equation}
In the limit as $\lambda \rightarrow 0$ (at fixed
$(\bar{t},\bar{r},\theta,\phi)$) the ``deSitter background'' becomes
irrelevant. The limiting metric is simply the Schwarzschild metric of
unit mass. The fact that the limit as $\lambda \rightarrow 0$ exists
can be attributed to the fact that the Schwarzschild black hole is
shrinking to zero in a manner where, in essence, nothing changes
except the overall scale.

The simultaneous existence of both of the above types of limits
charaterizes the type of one-parameter family of spacetimes $g_{\mu
  \nu} (\lambda)$ that we wish to consider. More precisely,
we wish to consider a one parameter family of solutions $g_{\mu\nu} (\lambda)$
satisfying the following properties:

\begin{itemize}

\item 
(i) Existence of the ``ordinary limit'': There exist coordinates $x^\alpha$
  such that $g_{\mu \nu}(\lambda, x^\alpha)$ is jointly smooth in
  $(\lambda, x^\alpha)$, at least for $r > \bar{R} \lambda$ for some
  constant $\bar{R}$, where $r \equiv \sqrt{\sum (x^i)^2}$
  ($i=1,2,3$). For all $\lambda$ and for $r > \bar{R} \lambda$,
  $g_{\mu\nu}(\lambda)$ is a vacuum solution of Einstein's
  equation. Furthermore, $g_{\mu \nu}(\lambda = 0, x^\alpha)$ is
  smooth in $x^\alpha$, including at $r= 0$, and, for $\lambda=0$, the
  curve $\gamma$ defined by $r= 0$ is timelike.

\item
(ii) Existence of the ``scaled limit'': For each $t_0$,
  we define $\bar{t} \equiv (t-t_0)/\lambda$, $\bar{x}^i \equiv
  x^i/\lambda$.  Then the metric $\bar{g}_{\bar{\mu}
  \bar{\nu}}(\lambda; t_0; \bar{x}^\alpha) \equiv \lambda^{-2}
  g_{\bar{\mu} \bar{\nu}}(\lambda; t_0; \bar{x}^\alpha)$ is jointly
  smooth in $(\lambda, t_0; \bar{x}^\alpha)$ for $\bar{r} \equiv
  r/\lambda > \bar{R}$.

\end{itemize}

The above two conditions must be supplemented by an additional ``uniformity requirement'', which can be explained as follows.
From the definitions of $\bar{g}_{\bar{\mu} \bar{\nu}}$ and $\bar{x}^\mu$, we can relate coordinate components of the barred metric in barred coordinates to coordinate components of the unbarred metric in corresponding unbarred coordinates,
\begin{equation}
\bar{g}_{\bar{\mu} \bar{\nu}}(\lambda; t_0; \bar{t},\bar{x}^i)
= g_{\mu \nu} (\lambda; t_0 + \lambda \bar{t}, \lambda \bar{x}^i)
\,.
\end{equation}
Now introduce new variables $\alpha \equiv r$ and $\beta \equiv \lambda/r
= 1/\bar{r}$, and view the metric components $g_{\mu\nu}(\lambda)$ as functions of $(\alpha, \beta, t, \theta, \phi)$, where $\theta$ and $\phi$ are defined in terms of $x^i$ by the usual formula for spherical polar angles.  We have
\begin{equation}
\bar{g}_{\bar{\mu} \bar{\nu}}(\alpha \beta, t_0; \bar{t}, 1/\beta, \theta,\phi) = 
g_{\mu \nu}(\alpha \beta, t=t_0 + \lambda \bar{t}; \alpha, \theta, \phi) \, .
\end{equation}
Then, by assumption (ii) we see that for $0 <
\beta < 1/\bar{R}$, $g_{\mu \nu}$ is smooth in $(\alpha, \beta)$ for all
$\alpha$ {\it including} $\alpha = 0$. By assumption (i),
we see that
for all $\alpha > 0$, $g_{\mu \nu}$ is smooth in $(\alpha, \beta)$ for $\beta <
1/\bar{R}$, {\it including} $\beta = 0$. Furthermore, for $\beta = 0$,
$g_{\mu \nu}$ is smooth in $\alpha$, {\it including} $\alpha = 0$.

We now impose the additional uniformity requirement on
our one-parameter family of spacetimes:

\begin{itemize}
\item
(iii) $g_{\mu \nu}$ is jointly smooth in $(\alpha, \beta)$ at $(0,0)$. 
\end{itemize}

We already know from our previous assumptions that $g_{\mu \nu}(\lambda; t_0, r, \theta, \phi)$ and its derivatives with
respect to $x^\alpha$ approach a limit if we let $\lambda \rightarrow
0$ at fixed $r$ and then let $r \rightarrow 0$. The uniformity
requirement implies that the same limits are attained whenever
$\lambda$ and $r$ both go to zero in any way such that $\lambda/r$
goes to zero.  

It has recently been proven in \cite{gralla-harte-wald} that an analog of the uniformity
requirement holds for electromagnetism in Minkowski spacetime in the
following sense: Consider a one-parameter family of charge-current
sources of the form $J^\mu(\lambda,t,x^i) =
\tilde{J}^\mu(\lambda,t,x^i/\lambda)$ where $ \tilde{J}^\mu$
is a smooth function of its arguments and $x^i = 0$ defines a
timelike worldline. Then the retarded solution, $F_{\mu\nu}(\lambda,x^\mu)$,
is a smooth function the variables $(\alpha, \beta, t, \theta, \phi)$ in a 
neigborhood of $(\alpha, \beta) = (0,0)$. In the gravitational case, we do not have 
a simple relationship between the metric and the stress-energy source, and in the nonlinear
regime, it would not make sense to formulate the uniformity condition in terms of
the behavior of the stress-energy. Consequently, we have formulated this condition in
terms of the behavior of the metric itself.

The uniformity requirement implies that the metric components can be approximated near $(\alpha,\beta) = (0,0)$ with a finite Taylor series in $\alpha$ and $\beta$,
\begin{equation}\label{FZ}
g_{\mu \nu}(\lambda;t,r,\theta,\phi) = \displaystyle \sum_{n=0}^{N}
\sum_{m=0}^{M} r^n \left(\frac{\lambda}{r}\right)^m
(a_{\mu \nu})_{nm}(t,\theta,\phi),
\end{equation}
where remainder terms have been dropped.  This gives a {\it far zone expansion}. Equivalently, we have
\begin{equation}
\bar{g}_{\bar{\mu} \bar{\nu}}(\lambda; t_0;
\bar{t},\bar{r},\theta,\phi) = \displaystyle \sum_{n=0}^{N}
\sum_{m=0}^{M} (\lambda \bar{r})^n \left( \frac{1}{\bar{r}} \right)^m
(a_{\mu \nu})_{nm}(t_0 + \lambda \bar{t},\theta,\phi) \, .
\end{equation}
Further Taylor expanding this formula with respect to the time variable yields
a {\it near zone expansion}.
Note that since we can express $\bar{g}_{\bar{\mu} \bar{\nu}}$ at $\lambda =0$
as a series in $1/\bar{r}$ as $\bar{r} \rightarrow \infty$ 
and since $\bar{g}_{\bar{\mu} \bar{\nu}}$ at $\lambda =0$
does not depend on $\bar{t}$, we see that
$\bar{g}_{\bar{\mu} \bar{\nu}} (\lambda =0)$ is a stationary,
asymptotically flat spacetime. 

The curve $\gamma$ to which our body shrinks as $\lambda \rightarrow 0$ (see
condition (i) above) can now be proven to be a geodesic of the metric
$g_{\mu\nu}(\lambda=0)$ as follows: Choose the coordinates $x^\alpha$ so that at $\lambda = 0$ they correspond to Fermi normal coordinates about the worldline $\gamma$. In particular, we have $g_{\mu\nu} = \eta_{\mu\nu}$ on $\gamma$ at $\lambda=0$.
It follows from \eqref{FZ} that near $\gamma$ (i.e.,
for small $r$) the metric $g_{\mu \nu}$ must take the form
\begin{equation}
 g_{\mu \nu} = \eta_{\mu \nu} + O(r) + \lambda \left(
 \frac{C_{\mu \nu}(t, \theta, \phi)}{r}+O(1)
 \right)+O(\lambda^2) \,
\end{equation}
Now, for $r>0$, the coefficient of $\lambda$, namely
\begin{equation}
h_{\mu \nu} = \frac{C_{\mu \nu}}{r}\ +O(1)
\end{equation}
must satisfy the vacuum linearized Einstein equation off of the
background spacetime $g_{\mu \nu} (\lambda = 0)$. However, since each
component of $h_{\mu \nu}$ is a locally $L^1$ function, it
follows immediately that $h_{\mu \nu}$ is well defined as a
distribution. It is not difficult to show that, as a distribution,
$h_{\mu \nu}$ satisfies the linearized Einstein equation with
source of the form $N_{\mu \nu}(t) \delta^{(3)} (x^i)$, where $N_{\mu
\nu}$ is given by a formula involving the limit as $r \rightarrow 0$
of the angular average of $C_{\mu \nu}$ and its first derivative.  The
linearized Bianchi identity then immediately implies that
$N_{\mu\nu}$ is of the form $M u_\mu u_\nu$ with $M$ constant, and that $\gamma$ is a geodesic for $M \neq 0$.

Our main interest, however, is not to rederive geodesic motion but to
find the leading order corrections to geodesic motion that arise from
finite mass and finite size effects. To define these corrections, we
need to have a notion of the ``location'' of the body to first order
in $\lambda$. This can be defined as follows: Since
$\bar{g}_{\bar{\mu} \bar{\nu}}(\lambda = 0)$ is an asymptotically flat
spacetime, its mass dipole moment can be set to zero (at all
$t_0$) as a gauge condition on the coordinates $\bar{x}^i$. The new
coordinates $\bar{x}^i$ then have the interpretation of being ``center
of mass coordinates'' for the spacetime $\bar{g}_{\bar{\mu}
  \bar{\nu}}(\lambda = 0)$.  In terms of our original coordinates
$x^\alpha$, the transformation to center of mass coordinates at all
$t_0$ corresponds to a coordinate transformation $x^\alpha \rightarrow
\hat{x}^\alpha$ of the form
\begin{equation}
\hat{x}^\alpha (t) = x^\alpha - \lambda A^\alpha (t, x^i)
+ O(\lambda^2) \, .
\end{equation}
To first order in $\lambda$, the world line defined by $\hat{x}^i = 0$
should correspond to the ``position'' of the body.  The first-order displacement from $\gamma$ in the original coordinates is then given simply by
\begin{equation}
Z^i(t) \equiv A^i(t,x^j=0) .
\end{equation}
The quantity $Z^i$ is most naturally interpreted as a ``deviation
vector field'' defined on $\gamma$.
Our goal is to derive relations (if any) that hold for $Z^i$ that are
independent of the choice of one-parameter family satisfying our
assumptions.

We now choose the $x^\alpha$ coordinates---previously chosen to agree
with Fermi normal coordinates on $\gamma$ at $\lambda=0$---to
correspond to the Lorenz/harmonic gauge to first order in $\lambda$. To order
$\lambda^2$, the leading order in $r$ terms in $g_{\alpha \beta}$ are,
\begin{equation}\label{eq:metric2}
\begin{split}
g_{\alpha \beta}(\lambda;t,x^i) & =  \eta_{\alpha \beta} + B_{\alpha i
  \beta j}(t)x^i x^j + O(r^3) \\ & \quad +  \lambda \left(
\frac{2M}{r}\delta_{\alpha \beta} + h^{\textrm{\tiny
    tail}}_{\alpha \beta}(t,0) + h^{\textrm{\tiny tail}}_{\alpha
  \beta i}(t,0)x^i + M \mathcal{R}_{\alpha \beta}(t) + O(r^2) \right) 
\\ & \quad +  \lambda^2 \biggl( \frac{M^2}{r^2} \left( -2t_\alpha t_\beta + 3
n_\alpha n_\beta \right) + \frac{2}{r^2} P_i(t) n^i \delta_{\alpha
  \beta} + \frac{1}{r^2} t_{(\alpha} S_{\beta) j}(t) n^j \\
& \qquad \qquad + \frac{1}{r}
K_{\alpha \beta}(t,\theta,\phi) + H_{\alpha \beta}(t,\theta,\phi) +
O(r) \biggr) + O(\lambda^3) 
\end{split}
\end{equation}
Here $B$ and $\mathcal{R}$ are expressions involving the curvature of
$g_{\mu \nu}(\lambda = 0)$ and we have introduced the ``unknown'' tensors
$K$ and $H$.  The quantities $P^i$ and $S_{\alpha \beta}$ turn out to be the mass dipole and spin of the ``near-zone'' background spacetime $\bar{g}_{\bar{\mu}\bar{\nu}}(\lambda=0)$. For simplicity, we have assumed no ``incoming radiation''. 
Hadamard expansion techniques and 2nd order perturbation theory were 
used to derive this expression.

Using the coordinate shift $x^\mu \rightarrow x^\mu - \lambda A^\mu$ to cancel
the mass dipole term, the above expression translates into the
following expression for the scaled metric
\begin{align}
\bar{g}_{\bar{\alpha} \bar{\beta}}(\hat{t}_0)  &= \eta_{\alpha \beta}
+  \frac{2M}{\bar{r}} \delta_{\alpha \beta} + \frac{M^2}{\bar{r}^2}
\left( -2t_\alpha t_\beta + 3 n_\alpha n_\beta \right) \nonumber + \frac{1}{\bar{r}^2} t_{(\alpha} S_{\beta) j} n^j + O \left(
\frac{1}{\bar{r}^3} \right) \nonumber \\
&\quad +  \lambda \left[
  h^{\textrm{\tiny tail}}_{\alpha \beta} + 2 A_{(\alpha , \beta)}
  + \frac{1}{r} K_{\alpha \beta} \nonumber + \frac{\bar{t}}{\bar{r}^2} t_{(\alpha} \dot{S}_{\beta) j} n^j 
+ O \left( \frac{1}{\bar{r}^2}
  \right) + \bar{t} \ O\left( \frac{1}{\bar{r}^3} \right)\right] \nonumber \\
&\quad + \lambda^2 \bigg[ B_{\alpha i \beta j} \bar{x}^i \bar{x}^j +
  h^{\textrm{\tiny tail}}_{\alpha \beta , \gamma}\bar{x}^\gamma +
  M \mathcal{R}_{\alpha \beta}(\bar{x}^i) + 2 B_{\alpha i \beta j}A^i
  \bar{x}^j + 2 A_{(\alpha , \beta) \gamma} \bar{x}^\gamma \nonumber \\
  &\quad + H_{\alpha
 \beta} + \frac{\bar{t}}{\bar{r}} \dot{K}_{\alpha \beta} + \frac{\bar{t}^2}{
\bar{r}^2} t_{(\alpha} \ddot{S}_{\beta) j} n^j + O\left( \frac{1}{\bar{r}} \right) + \bar{t} \ O \left(
  \frac{1}{\bar{r}^3} \right) + \bar{t}^2 \ O \left(
  \frac{1}{\bar{r}^3} \right) ] + O(\lambda^3) \bigg] \,\, .
\end{align}

The terms that are first order in $\lambda$ in this equation satisfy
the linearized vacuum Einstein equation about the background ``near
zone'' metric (i.e., the terms that are $0$th order in $\lambda$). 
From this equation, we find that $dS_{ij}/dt = 0$,
i.e., to lowest order, spin is parallelly propagated along $\gamma$.

The terms that are second order in $\lambda$ in this equation 
satisfy the linearized
Einstein equation about the background
``near zone'' metric
with source given by the second order Einstein tensor of the first
order terms. Extracting the $\ell = 1$, electric
parity, even-under-time-reversal part of this equation at $O(1/\bar{r}^2)$ and $O(\bar{t}/\bar{r}^3)$, we obtain (after considerable algebra!)
\begin{equation}
Z_{i,00} = \frac{1}{2M} S^{kl} R_{kl0i} - R_{0 j 0 i}Z^j - 
\left( h^{\textrm{\tiny tail}}_{i0,0} - \frac{1}{2} 
h^{\textrm{\tiny tail}}_{00,i} \right)\,\, .
\end{equation}
In other words, in the Lorenz gauge, the deviation vector field,
$Z^a$, on $\gamma$ that
describes the first order perturbation to the motion satisfies
\begin{equation}
u^c\nabla_c(u^b\nabla_b Z^a) = \frac{1}{2M} {R_{bcd}}^a S^{bc} u^d 
- {R_{bcd}}^au^b u^d Z^c - (g^{ab} + u^a u^b)(\nabla_d
h_{bc}^{\tiny \textrm{tail}}- \frac{1}{2} \nabla_b h_{cd}^{\tiny
\textrm{tail}})u^c u^d \,\, .
\label{EOM}
\end{equation}
Equation (\ref{EOM}) gives the desired leading order corrections to
motion along the geodesic $\gamma$. The first term on the right side
of this equation is the Papapetrou ``spin force'', which is the
leading order ``finite size'' correction. The second term is just the
usual right hand side of the geodesic deviation equation; it is not a
correction to geodesic motion but rather allows for the possibility
that the perturbation may make the body move along a different
geodesic. Finally, the last term describes the gravitational
self-force that we had sought to obtain, i.e., the corrections to the
motion caused by the body's self-field.  Equation \eqref{EOM} gives the correct description of motion when the metric perturbation is in the Lorenz gauge.  When the metric perturbation is expressed in a different gauge, the force will be different \cite{gralla-wald}.

Although we have now obtained the perturbative correction to geodesic
motion due to spin and self-force effects, at late times the small
corrections due to self-force effects should accumulate (e.g., during
an inspiral), and eventually the orbit should deviate significantly
from the original, unperturbed geodesic $\gamma$. When this happens,
it is clear our perturbative description in terms of a deviation
vector defined on $\gamma$ will not be accurate. Clearly, going to any
(finite) higher order in perturbation theory will not help (much).
However, if the mass and size of the body are sufficiently small, we
expect that its motion is well described {\it locally} as a
small perturbation of {\it some} geodesic. {\it Therefore, one
should obtain a good description of the motion by making
up (!) a ``self-consistent perturbative equation''} that satisfies
the following criteria: (1) It has a well posed initial
value formulation. (2) It has the same ``number of
degrees of freedom'' as the original system. (3) Its
solutions correspond closely to the solutions of the of the original
perturbation equation over a time interval where the perturbation
remains small. In some sense, such a self-consistent perturbative
equation would take into account the important (``secular'') higher
order perturbative effects (to all orders), but ignore other higher
order corrections.  Such equations are commonly considered in physics.
The MiSaTaQuWa equations appear to be a good candidate for
a self-consistent perturbative equation associated with our perturbative result.

In summary, we have analyzed the motion of a small body or black hole
in general relativity, assuming only the existence of a one-parameter
family of solutions satisfying assumptions (i), (ii), and (iii) above.
We showed that at lowest (``zeroth'') order, the motion of a
``small'' body is described by a geodesic, $\gamma$, of the ``background''
spaceetime. We then derived is a formula for the first order
deviation of the ``center of mass'' worldline of the body from $\gamma$.
The MiSaTaQuWa equations then
arise as (candidate) ``self-consistent
perturbative equations'' based on our first order perturbative 
result. Note that it is only at this stage that
``phoney'' linearized Einstein equations come into play.

We have recently applied this basic approach to the derivation
of self-force in electromagnetism \cite{gralla-harte-wald}, and have argued that the reduced
order form of the Abraham-Lorentz-Dirac equation provides an appropriate self-consistent
perturbative equation associated with our first order perturbative result (whereas the original Abraham-Lorentz-Dirac equation is excluded).
It should be possible to use this formalism to take higher order
corrections to the motion into account in a systematic way in both the gravitational and 
electromagnetic cases.

\bigskip

\noindent {\bf Acknowledgments}

This research was supported in part by NSF grants PHY04-56619 
and PHY08-54807 to the
University of Chicago.

\end{document}